\documentclass{iopart}
\usepackage{graphicx,graphics}
\begin{document}
\title[Effect of Transparency on the D-wave Josephson Junction]{Effect of Transparency on the Josephson Junction between D-wave Superconductors}
\author{Gholamreza Rashedi}
\address{Department of Physics, University of Isfahan, Isfahan,
Iran}\ead{rashedi@phys.ui.ac.ir}
\begin{abstract}
In this paper, a dc Josephson junction between two singlet
superconductors (d-wave and s-wave) with arbitrary reflection
coefficient has been investigated theoretically following the
famous paper [ Y. Tanaka and S. Kashiwaya 1996 \textit{Phys. Rev.
B} {\bf 53}, R11957]. For the case of High $T_c$ superconductors,
the $c$-axes are parallel to an interface with finite transparency
and their $ab$-planes have a mis-orientation. The effect of
transparency and mis-orientation on the currents is studied both
analytically and numerically. It is observed that, the current
phase relations are totally different from the case of ideal
transparent Josephson junctions between d-wave superconductors and
two s-wave superconductors. This apparatus can be used to
demonstrate d-wave order parameter in High $T_c$ superconductors.
\end{abstract}
\begin{figure}
\centering{
\resizebox{0.6\textwidth}{0.35\textheight}{\includegraphics{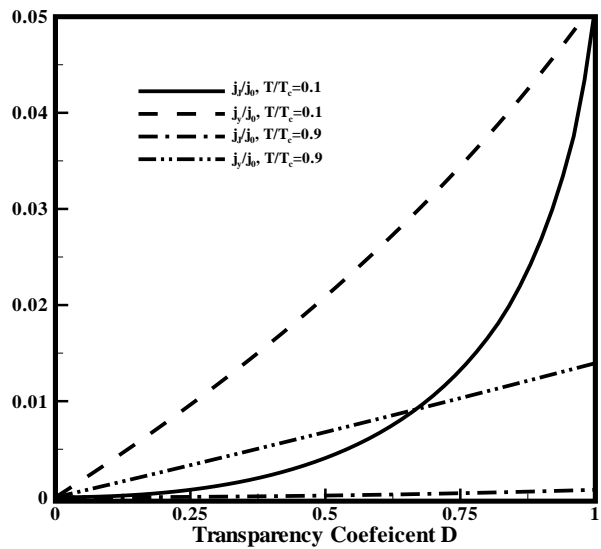}}}
 \caption{Critical Josephson and spontaneous currents versus transparency for
 $\alpha_l=0$, $\alpha_r=\frac{\pi}{4}$ and $T/T_C=0.1, 0.9$.} \label{pic5}
 \end{figure}

\pacs{74.50.+r,85.25.Cp,74.20.Rp,85.25.Dq,74.78.Bz} The quantum
transport through the junctions between high $T_c$ unconventional
superconductors has attracted much attention in recent years,
particularly in view of various recently discovered copper oxide
compounds. It is well-known that the isotope effect for these
superconductors is rejected or weak. The isotope effect is the
most important reason for the phonon participation in the
superconductivity mechanism. Thus we should search for another
pairing mechanism for high-$T_c$ superconductors. In spite of
enormous efforts made to understand the physical mechanisms of
pairing in various high-$T_c$ superconductors, the situation still
remains unclear. A central role in understanding of this type of
superconductivity relates to the symmetry of the order parameter.
Soon after the first Bednords and Muller \cite{J} discovery, the
symmetry of the $d_{x^2-y^2}$-form was suggested for high-$T_c$
superconducting materials. It was considered that the formation of
midgap states (MGS) at surfaces and interfaces of $d$-wave
superconductors affects the current transport properties of
junctions involving $d$-wave superconductors \cite{C}. It is found
that in spite of $s$-wave superconductors for $d$-wave
superconductors we can observe midgap states. This phenomenon is
due to the changing the sign of order parameter in $d$-wave
superconductors. On the other hand, measurements of direct
Josephson current yield valuable information on the symmetry of
the order parameter which is essential for understanding the
mechanisms of superconductivity in these complex materials. Phase
interference experiments definitely suggest the presence of
$d$-wave symmetry of the order parameter in high-$T_c$
superconductors \cite{D,BRO}. The properties of spontaneous
currents generated at surfaces and interfaces of $d$-wave
superconductors using the self-consistent quasiclassical
Eilenberger equations are investigated \cite{M}. The effect of
transparency and mis-orientation on the current as the main
purpose of this paper. In this paper, we have solved the nonlocal
Eilenberger equations \cite{G} supplemented with Zaitsev boundary
conditions \cite{AZ} and obtained the corresponding Green
functions analytically. These boundary conditions have been
rewritten in \cite{BE,RA}. Then, using the Green functions, we
compute the spontaneous and Josephson currents. The system which
is investigated in this paper, consists of two $d$-wave
superconductors connecting through an interface. The interface has
the finite transparency for quasiparticles and two superconductors
have a mis-orientation with each other. Quasiclassical theory of
superconductivity is based on the nonlocal Eilenberger's equations
for the quasiclassical matrix propagator.
\begin{figure}[ht]\centering
{\resizebox{0.6\textwidth}{0.35\textheight}{\includegraphics{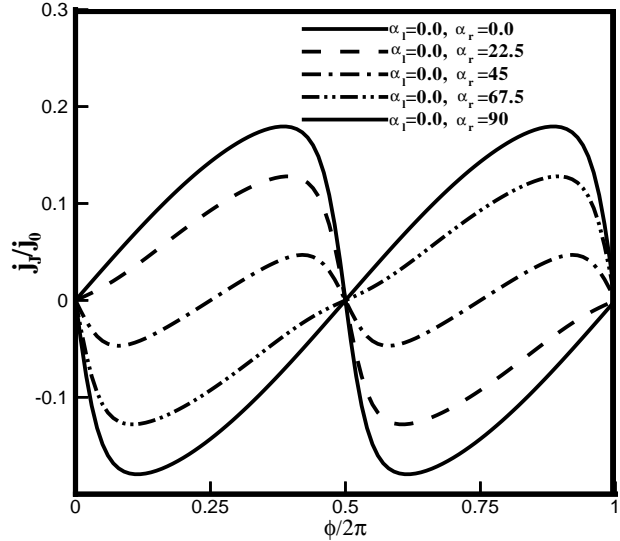}}}
\caption{Josephson current through the interface versus phase for
$D=0.99$, $\alpha_l=0$, $T/T_c=0.1$, $j_{0}=4\protect\pi e
N(0)v_{F}T_{c}$.}\label{pic1}
\end{figure}
In the case of a clean (ballistic) singlet anisotropically paired
superconductor, the equations for the $\xi$-integrated Green
functions reduce to the following $2\times2$ matrix form \cite{G}
\begin{equation}
{\bf v}_{F}\cdot\frac{\partial}{\partial{\bf
r}}\widehat{G}_{\omega}({\bf v}_{F},{\bf r})
+[\omega\widehat{\tau}_{3}+\widehat{\Delta}({\bf v}_{F},{\bf r}),
\widehat{G}_{\omega}({\bf v}_{F},{\bf r})]=0 \label{EIL}
 \end{equation}
 where
 \begin{equation}
 \widehat{\Delta }=\left(
 \begin{array}{cc}
      0 & \Delta \\
      \Delta ^{\dagger } & 0
      \end{array}
      \right) ,\quad \widehat{G}_{\omega }({\bf v}_{F},{\bf r})=\left(
      \begin{array}{cc}
      g_{\omega } & f_{\omega } \\
      f_{\omega }^{\dagger } & -g_{\omega }
      \end{array}
      \right) \label{DEL}
      \end{equation}
${\Delta}$ is the superconducting order parameter,
$\widehat{\tau}_{3}$is the Pauli matrix, and $\widehat{G}_{\omega
}({\bf v}_{F},{\bf r})$ is the matrix Green function which depends
on the electron velocity on the Fermi surface ${\bf v}_{F}$, the
coordinate ${\bf r}$ and the Matsubara frequency
${\omega}=(2n+1)\pi T$ ,with $n$ and $T$ being an integer number
and temperature respectively. We also need to satisfy the
normalization condition, $ g_{\omega }={\sqrt {1-f_{\omega
}f_{\omega }^{\dagger }}}$, with $f_{\omega}^{\dagger}$ being
time-reversal counterpart of $f_{\omega}$. In general, $\Delta $
 depends on the direction of ${\bf v} _{F}$ and is determined by the self-consistency equation
    \begin{equation}
     \Delta ({\bf v}_{F},{\bf r})=2\pi N(0)T\sum\limits_{\omega >0}
     \left< V( {\bf v}_{F},{\bf v}_{F}^{\prime })f_{\omega }({\bf
     v}_{F}^{\prime }, {\bf r}) \right>_{{\bf v}_{F}^{^{\prime
     }}}\label{SEL}
     \end{equation}
where $V({\bf v}_{F},{\bf v}_{F}^{\prime })$ is the interaction
potential. Solution of the matrix equation (\ref{EIL}) together
with self-consistency equation (\ref{SEL}) and normalization
condition determines the current density ${\bf j(r)}$ in the
system
     \begin{equation}
      {\bf j(r)}=-4\pi ieN(0)T\sum\limits_{\omega }\left\langle {\bf v}%
     _{F}g_{\omega }({\bf v}_{F},{\bf r})\right\rangle _{{\bf
     v}_{F}}\label{CUR}.
      \end{equation}
In two dimensions, $N(0)=m/ 2\pi$ is the 2D density of states and
$\left<... \right>=\int\limits_{0}^{2\pi } (d\theta / 2\pi )...$
is the averaging over directions of 2D vector ${\bf v }_{F}$. A
symmetric configuration is considered and the equations are
changed to a two-dimensional problem. Thus the spatial
distributions of $\Delta({\bf r})$ and ${\bf j}({\bf r})$ depend
only on the coordinates in the $ab$-plane, and Eilenberger
equations (\ref{EIL}) reduce to a simpler problem. In the present
paper, it is assumed that transparency of the interface to be
finite $0{\leq}D{\leq}1$. Also, the equations (\ref{EIL}) for
Green function $\widehat{G}_{\omega}({\bf v}_{F},{\bf r})$ have to
be supplemented by the suitable boundary conditions
\cite{AZ,BE,RA}. These boundary conditions couple the transmitted
and reflected trajectories \cite{BE}. So, because of finite
reflection at the interface, it is impossible to impose the
continuity of the Green functions along a trajectory that crosses
over the interface, as done for the ideal transparent system
\cite{KO}.  Instead, the Zaitsev boundary conditions \cite{AZ}
should be exerted as follows
\begin{equation}
\widehat{d}^{~l}=\widehat{d}^{~r}\equiv \widehat{d}\label{Z1}
\end{equation}
\begin{figure}[ht]\centering{
\resizebox{0.6\textwidth}{0.35\textheight}{\includegraphics{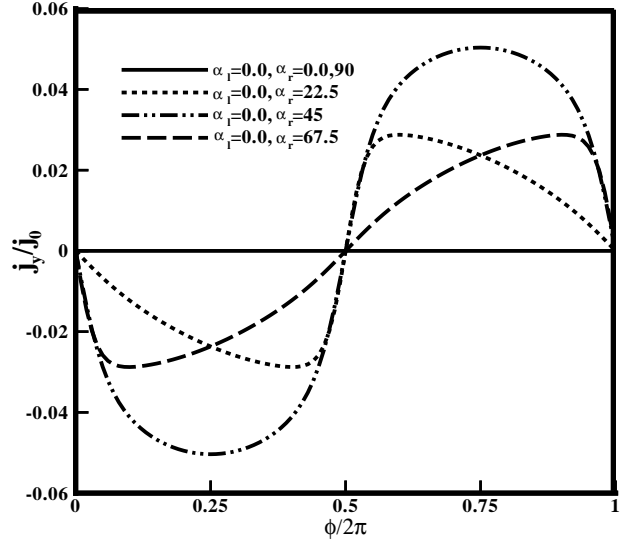}}}
\caption{Parallel,spontaneous current $j_{y}(0)$ versus phase $\phi$
for $T/T_c=0.1$, $\alpha_l=0$, $D=0.99$ and different $\alpha_r$s.}
\label{pic2}
\end{figure}
\begin{equation}
\left(\frac{D}{2-D}\right)\left[( 1+{\frac{\widehat{d}}{2}})
\widehat{s}^{~r} , \ \widehat{s}^{~l}\right] =\widehat{d}\left(\
\widehat{s}^{~l}\right)^{2}\label{Z2}
\end{equation}
where
\begin{equation}
\widehat{s}^{~r}=\widehat{G}_{\omega }^{~r}({\bf
v}_{F},x{=}0)+\widehat{G}_{\omega }^{~r}({\bf v}_{F}^{\prime
},x{=}0) \label{SYM}
\end{equation}
and
\begin{equation}
 \widehat{d}^{~r}=\widehat{G}_{\omega }^{~r}({\bf
v}_{F},x{=}0)-\widehat{G}_{\omega }^{~r}({\bf v}_{F}^{\prime
},x{=}0) \label{ANS}
\end{equation}
 with ${\bf v}_{F}^{\prime }$ being the reflection of ${\bf
v}_{F}$with respect to the boundary and $D$ is transparency
coefficient of the interface that can be momentum dependent
generally.
 Similar relations also hold for $\widehat{s}^{~l}$  and
$\widehat{d}^{~l}$. In our calculations, $ \Delta_{l,r}({\bf
v}_F)=\Delta_0(T){\cos}(2(\theta-\alpha_{l,r}))$ are used for the
left and right side of the interface respectively with
$d_{x^2-y^2}$ order parameters symmetry and $\alpha_{l,r}$ being
orientations with respect to the interface. It is remarkable that
a step function for the spatial dependence for the phase of order
parameters are considered as $\Delta_{l,r}(x)=\Delta_{l,r}\exp{\pm
i\phi/2}$ respectively and for simplicity the self-consistency of
problem is ignored. For all quasiparticle trajectories the Green
functions satisfy the boundary conditions both in the right and
left bulks and at the interface.
\begin{figure}
\centering{
\resizebox{0.6\textwidth}{0.35\textheight}{\includegraphics{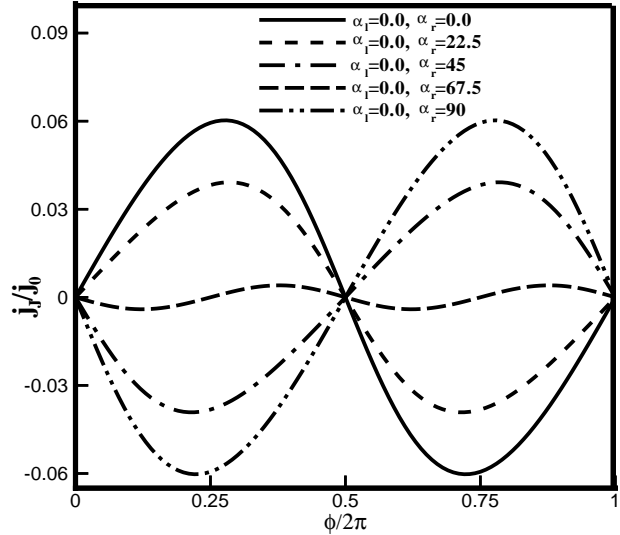}}}
\caption{Josephson current through the interface versus phase for
$D=0.5$, $\alpha_l=0$, $T/T_c=0.1$ and different $\alpha_r$s.}
\label{pic3}
\end{figure}
We have solved this problem and obtained the Green functions of this
system analytically. Using our solution of Eilenberger equations and
symmetries of singlet order parameters, we obtain the following
expression for the current density (\ref{CUR}) exactly at the
interface
\begin{equation}
     {\bf j}(0)= j_{0}\frac{T}{T_{c}}\sum\limits_{\omega
     >0}\left\langle \frac{\widehat{{\bf
     v}}\eta\Delta_l\Delta_r
     \sin {\phi }}{\omega^{2}+\Delta_l\Delta_r
     \cos {\phi }+(\frac{2-D}{D})\Omega_l\Omega_r}\right\rangle _{\widehat{{\bf
     v}}}\label{DENS}
\end{equation}
where, $\Omega_{l,r}=\sqrt{{\omega}^{2}+\Delta_{l,r} ^{2}}$, with
$\omega$ being the Matsubara frequency, $\widehat{{\bf v}}={\bf
v}_{F}/v_{F}$ is the unit vector, $\eta=sign(v_{x})$, $\phi$ is the
phase difference between two bulks and $j_{0}=4\protect\pi e
N(0)v_{F}T_{c}$.\\ For the junction between two spin-singlet s-wave
superconductors, with the isotropic order parameters in the momentum
space $\Delta({\bf v}_{F})=\Delta$, at the junction we reproduce the
current density as follows:
 \begin{equation}
 {\bf j}(0)=\frac{j_{0}\Delta}{4T_{c}}\frac{D \sin{\phi}}{\sqrt{1-D
 {(\sin \frac{\phi }{2})}^{2}}}\tanh\frac{\Delta}{2T}\sqrt{1-D
 {(\sin \frac{\phi }{2})}^{2}}.\label{DENSW1}
 \end{equation}
Where, $D$ is assumed to be independent of momentum direction and
isotropic in the momentum space. This formulate coincides with the
results of \cite{TK1,TK2,TK3,FUR} and for ideal transparent point
contact with results of \cite{AMB,KO} . At the interface and near
the critical and zero temperatures we have
\begin{equation}
 {\bf j}(T \rightarrow T_c)=\frac{j_{0}D
\sin{\phi}}{8}\left(\frac{\Delta(T_c)}{T_{c}}\right)^{2}\label{DENSW2}
\end{equation}
and
\begin{equation}
 {\bf j}(T \rightarrow 0)=\frac{j_{0}\Delta(0)}{4T_{c}}\frac{D \sin{\phi}}{\sqrt{1-D
 {(\sin \frac{\phi }{2})}^{2}}}\label{DENSW3}
\end{equation}
respectively. On the other hand, for the small values of
transparency of interface between d-waves, the current expression
is the following
\begin{equation}
{\bf j}(0)= j_{0}\frac{T}{T_{c}}\sum\limits_{\omega
>0}\left\langle\frac{ \widehat{{\bf
v}} \eta\Delta_l\Delta_r D
\sin{\phi}}{2\Omega_l\Omega_r}\right\rangle_{\widehat{{\bf
v}}}\label{LOWT}
\end{equation}
that is the same as one of the relations in \cite{YZ}. As it is
clear from (\ref{DENS}), for high values of transparency, currents
are nonlinear functions of the transparency, $D$, because of
$(\frac{2-D}{D})$ term in (\ref{DENS}). While for the low values of
transparency currents are exactly linear functions of transparency
(\ref{LOWT}). On the other hand, when nonlinearity is removed, the
deviation from the sinusoidal form disappears. It is well-known that
the current in this system for the low temperatures does not have
sinusoidal form, while near the critical temperature it tends to the
sinus\cite{KO}. Also, this system at the low transparency, has a
current-phase dependence similar to the high temperature. For the
Josephson junction between d-wave superconductors, in addition to
the Josephson current another term of current is observed, called
the spontaneous current.
\begin{figure}
\centering{
\resizebox{0.6\textwidth}{0.35\textheight}{\includegraphics{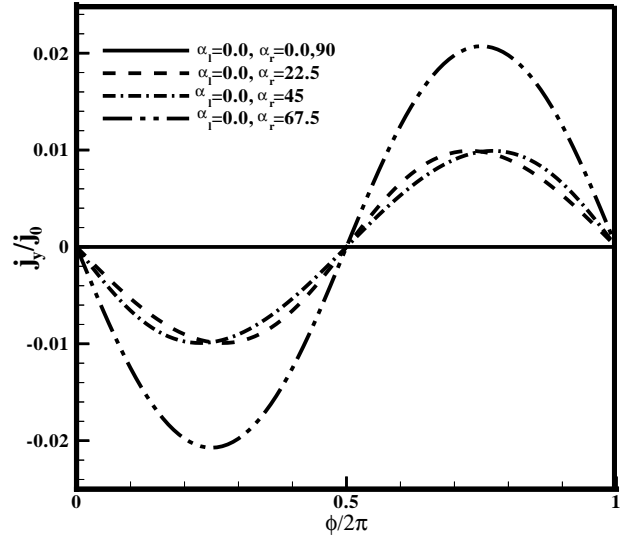}}}
 \caption{Parallel,spontaneous current $j_{y}(0)$ versus phase $\phi$
 for $T/T_c=0.1$, $\alpha_l=0$, $D=0.5$ and different $\alpha_r$s.} \label{pic4}
 \end{figure}
The spontaneous current that is parallel to the interface, is
plotted as a function of mis-orientation and transparency in
(Fig.\ref{pic2}) and (Fig.\ref{pic4}). The critical value of the
spontaneous current is plotted in terms of transparency in
(Fig.\ref{pic5}). The Josephson and spontaneous currents are plotted
as the functions of $\phi$ for different mis-orientations at
$T/T_c=0.1$, $D=0.99$ (Fig.\ref{pic1}), (Fig.\ref{pic3}) and for a
low value of transparency $D=0.5$ in (Fig.\ref{pic2}) and
(Fig.\ref{pic4}). By decreasing the transparency, the currents
decrease (Fig.\ref{pic5}). For high values of transparency the
currents are not sinusoidal function of phase, but by decreasing the
transparency they tend to the sinus as is observed in relations
(\ref{DENS}), (\ref{LOWT}). The period of current-phase plot is
interesting. We fixed the left orientation at $\alpha_l=0$, by
increasing the right orientation from $0$ to $\frac{\pi}{4}$, period
of the current-phase graphs varies from $2\pi$ to $\pi$ then, if the
right orientation changes to $\frac{\pi}{2}$, this period returns to
$2\pi$ (Fig.\ref{pic1}) and (Fig.\ref{pic2}). The critical values of
Josephson current in terms of mis-orientation are periodic function
of left and right mis-orientations. They are periodic function with
the period of $\pi$, because $\cos(2(x+\pi))=\cos(2x)$. For high
values of temperature, $T_c-T{\ll}T_c$, the Josephson current has
been obtained analytically and precisely as
\begin{equation}
 {\bf j}_J(0)=j_{0}\frac{4{\pi}(1-\frac{T}{T_c})}{315{\zeta(3)}}D\sin(\phi) [15\cos(2\alpha_l-2\alpha_r)
-\cos(2\alpha_l+2\alpha_r)]{\bf \hat{i}}\label{TCJ}
 \end{equation}
 and for the spontaneous current we have
 \begin{equation}
 {\bf j}_y(0)=\frac{j_{0}16{\pi}(1-\frac{T}{T_c})}{315{\zeta(3)}}D\sin \phi\sin(2\alpha_l+2\alpha_r)\label{TCS}
 \end{equation}
 where $\zeta(n)$ is the Riemann zeta function. For the $d$-wave order parameters at
 high temperatures, $T_c-T{\ll}T_c$, an analytical relation
$\Delta(T)=\sqrt{\frac{64\pi^{2}{T_c}^2}{21\zeta(3)}(1-\frac{T}{T_c})}$,
has been used. It is observed that close to the $T_c$, the currents
are exactly sinusoidal function of phase, quadratic function of
order parameter (linear dependence of $(T_c-T)$) and the linear
function of transparency, (Fig.\ref{pic5}), (\ref{TCJ}) and
(\ref{TCS}). The currents dependence on the orientations $\alpha_l$
and $\alpha_r$ are trigonometric functions and consequently periodic
with the period of $\pi$ generally. It is obtained that, in addition
to the mis-orientation $|\alpha_l-\alpha_r|$, the orientation with
respect to the interface is important, specially for the spontaneous
current because of the terms including $|\alpha_l+\alpha_r|$ in the
currents expressions.The spontaneous current for the high values of
transparency is an abnormal function of the phase, but for the small
values of transparency, it gets the sinus form (Fig.\ref{pic2}) and
(Fig.\ref{pic4}).\\
In conclusion, the effect of mis-orientation and transparency on
the currents through the Josephson junction in the \textbf{D-I-D}
system as well as in the \textbf{S-I-S} as an example, are
investigated. The authors of \cite{AP}, argued that, ${\bf
j}(\phi+\pi)=-{\bf j}(\phi)$, from my Green function is not
generally correct, because of $\cos(\phi)$ term in the denominator,
but for the high values of the temperature and small values of
transparency it will be correct.  For the fixed left orientation,
the current is a periodic function of the right orientation and the
vise versa. In addition to the mis-orientation,
$|\alpha_l-\alpha_r|$, the angle with the interface plays a role in
the current, because ${\bf j}(\alpha_l,\alpha_l)\neq{\bf
j}(\alpha_l+\beta,\alpha_l+\beta)$. This case is returned to the
scattering from the interface. The effect of transparency on the
currents, as in equations (\ref{LOWT}), (\ref{TCJ}), (\ref{TCS}) and
my numerical calculations (Fig.\ref{pic5}), for the high values of
temperature and low values of transparency is linear. But for high
values of transparency and the low temperatures the effect of
transparency is nonlinear (\ref{DENS}). Thus nonlinearity of
currents in terms of transparency ${\bf j}(D)$ is coupled with the
nonsinusoidal form of currents versus phase difference between order
parameters ${\bf j}(\phi)$. These currents dependence on the
transparency, phase difference and mis-orientation can be applied to
show the $d-$wave pairing symmetry for the high $T_c$
superconductors. As the most important result it is obtained that
${\bf j}(\alpha_l,\alpha_r,\phi+\pi)={\bf
j}(\alpha_l,\alpha_{r}+\frac{\pi}{2},\phi)$,  that means changing
the sign of order parameter on the Fermi surface changes the zero
junction to the $\pi$ junction.

\end{document}